# Blind user detection in doubly-dispersive DS/CDMA fading channels

Stefano Buzzi, *Senior Member, IEEE,* Luca Venturino, *Member, IEEE,* Alessio Zappone, *Student Member, IEEE,*
Antonio De Maio, *Senior Member, IEEE*

*Abstract*—In this work, we consider the problem of detecting the presence of a new user in a direct-sequence/code-division-multiple-access (DS/CDMA) system with a doubly-dispersive fading channel, and we propose a novel blind detection strategy which only requires knowledge of the spreading code of the user to be detected, but no prior information as to the time-varying channel impulse response and the structure of the multiaccess interference. The proposed detector has a bounded constant false alarm rate (CFAR) under the design assumptions, while providing satisfactory detection performance even in the presence of strong cochannel interference and high user mobility.

## I. INTRODUCTION

Code-division multiple-access (CDMA) is widely employed in third-generation (3G) cellular standards, and is a strong candidate for the definition of the air-interface of future wireless networks [1]–[3]. In this paper, we investigate the problem of detecting the presence of a new active user which naturally arises in the design of CDMA-based systems. Indeed, it is known that hidden (i.e., undetected) active terminals can cause a substantial degradation of the bit-error-rate performance even if the optimum maximum-likelihood multiuser receiver is employed [4]. If users enter and/or leave the channel at random epochs due to the bursty nature of the data transmission, detecting and exploiting these changes in the user population is crucial to achieve high spectral efficiencies and to guarantee a minimum quality of service to the set of active users. The problem of user acquisition may also be reframed in the context of cell-search and handoff procedures [5], wherein mobiles have to detect strong surrounding base stations and have to synchronize with their unique signatures transmitted on the broadcast channel, and in the context of neighbor discovery [6], [7], wherein mobiles sense the surrounding environment to identify other nodes.

First works on the detection of a spread-spectrum signal embedded in a mixture of multiaccess interference and thermal noise have focused on synchronous CDMA systems operating over non-fading or frequency-flat fading channels. In [8] a two-stage procedure is proposed, wherein the first stage resorts to linear or non-linear filtering to remove the multiple-access interference, while the second stage performs detection. In [9] a generalized likelihood detector is proposed, while a detection procedure based on the generalized cumulative sum test is presented in [10]. Finally, [11] proposes an innovative data detection structure based on the random set theory: despite its good performance, this approach has a prohibitive computational complexity. The work in [12], instead, discuss strategies for user detection in asynchronous DS/CDMA systems with a frequency-selective, slow fading channel: the proposed procedures are based on the application of the generalized likelihood ratio test (GLRT).

All these previous works critically rely on the assumption that the channel conditions remain constant for many symbol intervals. However, this assumption may not be met in practice since user mobility and terrain changes may induce rapid variations in the channel response, which become more pronounced as the carrier frequency increases [13]–[15]. The problem of achieving reliable communication over rapidly time-varying channels naturally arises for example in many popular data transmission standards (e.g., W-CDMA, UMTS, DVB-T and IEEE802.16), and in vehicular networks that have become more and more popular lately. Hence, it is of interest to develop improved algorithms that are capable of operating in this more hostile environment. Many papers tackling the problem of reliable communication over doubly-dispersive channels have appeared in the literature, [16]–[18] just to cite the most recent ones. However, the majority of these previous studies focus on orthogonal frequency division multiplexing (OFDM) systems, and do not address the problem of new user detection.

In this work, we consider a DS/CDMA system operating over a doubly-dispersive fading channel, and we assume that neither pilot symbol nor secondary data are available to estimate the channel impulse response and the structure of the cochannel interference. Under this challenging scenario, we first show that the GLRT approach followed in [12] cannot be employed anymore. Instead, we resort to the so-called *method of sieves*, a valuable statistical tool reported in [19], [20], and we derive a novel *modified* GLRT (MGLRT) which is fully-blind (i.e., it only requires knowledge of the spreading code of the user to be detected) and has a bounded constant false alarm rate (CFAR) under the design assumptions (i.e., the test statistic, although not enjoying itself the CFAR property, is upper-bounded by a CFAR test). Simulation results indicate that the proposed MGLRT is robust to the presence of strong cochannel interference and performs well even when the user to be detected is not active during the entire processing interval.

The remainder of the paper is organized as follows. In Section II the system model is presented. Section III contains the synthesis of the new detector. Section IV contains the numerical results, while concluding remarks are given in Section V.

*Notation:* In the following, $\boldsymbol{X}^T$, $\boldsymbol{X}^\dagger$, $\boldsymbol{X}^+$ and $\text{rank}\{\boldsymbol{X}\}$ denote

S. Buzzi, L. Venturino, and A. Zappone are with the DAEIMI, Università degli Studi di Cassino, Cassino (FR), ITALY. Phone: +39-0776-2993737, +39-0776-2994934, +39-0776-2994344, Fax: +39-0776-2993987, E-mail: buzzi@unicas.it, l.venturino@unicas.it, alessio.zappone@unicas.it.

A. De Maio is with the DIET, Università degli Studi di Napoli "Federico II", Napoli, ITALY. Phone: +39-081-7683147, Fax: +39-081-7683149, E-mail: ademaio@unina.it.

the transpose, the Hermitian transpose, the Moore-Penrose pseudoinverse, and the rank of a matrix $\boldsymbol{X} \in \mathbb{C}^{m \times n}$, respectively. tr$\{\boldsymbol{A}\}$, $|\boldsymbol{A}|$ and $|\boldsymbol{A}|_p$ denote the trace, the determinant, and the product of the positive eigenvalues of a square matrix $\boldsymbol{A}$, respectively. $\boldsymbol{I}_m$ and $\boldsymbol{O}_{n,m}$ indicate the identity matrix of order $m$, and an $n \times m$-dimensional matrix with all-zero entries, respectively.

## II. SYSTEM MODEL

Consider a DS/CDMA system with $K$ asynchronous users and denote by $s_k(t) = \sum_{n=0}^{N-1} \beta_k(n) \psi_{\text{tx}}(t - nT_c)$ the spreading signature assigned to user $k$, wherein $N$ is the processing gain, $T_c$ is the chip interval, $\boldsymbol{\beta}_k = [\beta_k(0), \ldots, \beta_k(N-1)]^T \in \mathbb{C}^N$ is the spreading code, and $\psi_{\text{tx}}(t)$ is a unit-energy pulse waveform that is non-zero in $[0, PT_c)$ with $P$ a positive integer. The complex envelope of the received signal can be written as

$$y(t) = \sum_{k=0}^{K-1} \sum_{q=-\infty}^{+\infty} b_k(q) \int_{-\infty}^{+\infty} s_k(t - \tau - \tau_k - qT_b) \times c_k(t, \tau) \mathrm{d}\tau + w(t). \quad (1)$$

In (1), $\tau_k \geq 0$ is the transmission delay of the $k$-th user; $b_k(q) \in \mathcal{B}_k$ is the data symbol transmitted by the $k$-th user during the $q$-th symbol interval of length $T_b = NT_c$, where $\mathcal{B}_k$ is an arbitrary (possibly user dependent) finite constellation with average unit energy; $c_k(t, \tau)$ is the equivalent time-variant channel response of the $k$-th user (accounting for the transmit signal power, the path loss and the multipath fading); finally, $w(t)$ is the additive noise, modeled as a circularly-symmetric, zero-mean white Gaussian process with power spectral density $\mathcal{N}_0$.

The received signal $y(t)$ is first sent to a linear filter matched to $\psi_{\text{tx}}(t)$. Letting $T_m$ be the maximum multipath delay spread of the channel, $\psi(t) = \int_{-\infty}^{+\infty} \psi_{\text{tx}}(u) \psi_{\text{tx}}^*(u - t + PT_c) \mathrm{d}u$, and assuming at the design stage that $c_k(t, \tau)$ remains constant for about $T_b + T_m + PT_c$ seconds, the output of the receive filter can be written as

$$r(t) = \int_{-\infty}^{+\infty} y(u) \psi_{\text{tx}}^*(u - t + PT_c) \mathrm{d}u$$
$$= \sum_{k=0}^{K-1} \sum_{q=-\infty}^{+\infty} b_k(q) \sum_{n=0}^{N-1} \beta_k(n) g_k(t - nT_c; q) + n(t), \quad (2)$$

wherein $n(t)$ is the filtered noise process, and we have shoved the channel-dependent quantities of the $k$-th user into the function

$$g_k(t; q) = \int_{-\infty}^{+\infty} \psi(t - \tau - \tau_k - qT_b) c_k(qT_b + \tau_k, \tau) \mathrm{d}\tau. \quad (3)$$

Notice that $g_k(t; q)$ has compact support in $[qT_b + \tau_k, qT_b + \tau_k + T_m + 2PT_c] \subseteq [qT_b, (q+1)T_b + 2PT_c)$, where the inclusion on the right hand side stems from the fairly reasonable assumption[1] $\tau_k + T_m < T_b$.

After sampling $r(t)$ at rate $T_c/M$, with $M$ a positive integer representing the number of samples per chip, we stack the complex-valued samples corresponding to the interval $[qT_b, (q+L)T_b]$, with $L \geq 2$, into the $LNM$-dimensional data vector $\boldsymbol{r}(q)$. Skipping the

[1]This assumption can be easily relaxed as shown in [21], [22].

analytical details, $\boldsymbol{r}(q)$ can be expressed as

$$\boldsymbol{r}(q) = \sum_{k=0}^{K-1} \sum_{\ell=-2}^{L-1} b_k(q+\ell) \boldsymbol{C}_{k,\ell} \boldsymbol{g}_k(q+\ell) + \boldsymbol{n}(q), \quad (4)$$

wherein $\boldsymbol{g}_k(q) \in \mathbb{C}^{(N+2P)M}$ is a channel-dependent vector which contains the samples $g_k(qT_b + nT_c/M; q)$ for $n = 0, \ldots, (N+2P)M - 1$; $\boldsymbol{n}(q) \in \mathbb{C}^{LNM}$ is a complex circularly-symmetric Gaussian vector with zero-mean and full-rank covariance matrix $\boldsymbol{R_n} \in \mathbb{C}^{LNM \times LNM}$, with $(\boldsymbol{R_n})_{i,j} = \mathcal{N}_0 \psi((i-j)T_c/M + PT_c)$; $\{\boldsymbol{C}_{k,\ell} \in \mathbb{C}^{LNM \times (N+2P)M}, \ell = -2, \ldots, L-1\}$ are code-dependent matrices which contain suitable shifts of the spreading vector $\boldsymbol{\beta}_k$. Indeed, let $\boldsymbol{A}_k \in \mathbb{C}^{(LN+2P-1)M \times (N+2P)M}$ be the matrix

$$\boldsymbol{A}_k = \begin{bmatrix} \boldsymbol{\beta}_k(0) & 0 & \ddots & 0 \\ \vdots & \boldsymbol{\beta}_k(0) & \ddots & \vdots \\ \vdots & \vdots & \ddots & 0 \\ \boldsymbol{\beta}_k(N-1) & \vdots & \ddots & \vdots \\ 0 & \boldsymbol{\beta}_k(N-1) & \ddots & 0 \\ \vdots & \vdots & \ddots & \boldsymbol{\beta}_k(0) \\ 0 & 0 & \ddots & \vdots \\ \vdots & \vdots & \ddots & \vdots \\ 0 & 0 & \ddots & \boldsymbol{\beta}_k(N-1) \\ \vdots & \vdots & \ddots & \vdots \end{bmatrix} \otimes \boldsymbol{I}_M, \quad (5)$$

the matrix $\boldsymbol{C}_{k,0}$ is obtained by taking the first $LNM$ rows of $\boldsymbol{A}_k$, while the matrices $\{\boldsymbol{C}_{k,\ell}, l \neq 0\}$ are obtained by taking suitable sub-blocks of $\boldsymbol{C}_{k,0}$ [12], [21]. It is worth pointing out that the discrete-time signal representation (4) is very powerful since it isolates the known code-dependant matrices from the unknown data symbols and the unknown channel-dependent vectors.

## III. DETECTOR DESIGN

In order to cope with the time-varying nature of the channel, $Q$ consecutive data vectors are jointly processed by the detector. Thus, the problem of detecting the presence of a new user, say user 0, in the received data multiplex can be formulated in terms of the following binary hypothesis test:

$$\begin{cases} H_0 : \boldsymbol{r}(q) = \boldsymbol{w}(q), & q = 1, \ldots, Q, \\ H_1 : \boldsymbol{r}(q) = b_0(q) \boldsymbol{C}_{0,0} \boldsymbol{g}_0(q) + \boldsymbol{z}(q), & q = 1, \ldots, Q, \end{cases} \quad (6)$$

where

$$\boldsymbol{w}(q) = \sum_{k=1}^{K-1} \sum_{\ell=-2}^{L-1} b_k(q+\ell) \boldsymbol{C}_{k,\ell} \boldsymbol{g}_k(q+\ell) + \boldsymbol{n}(q),$$

$$\boldsymbol{z}(q) = \boldsymbol{w}(q) + \sum_{\substack{\ell=-2 \\ \ell \neq 0}}^{L-1} b_0(q+\ell) \boldsymbol{C}_{0,\ell} \boldsymbol{g}_0(q+\ell).$$

Henceforth we assume $Q \geq LNM$, which implies that the data matrix $\boldsymbol{R} = [\boldsymbol{r}(1), \ldots, \boldsymbol{r}(Q)] \in \mathbb{C}^{LNM \times Q}$ is full row rank with probability one under both hypotheses. Also, under $H_0$, we model $w(1), \ldots, w(Q)$ as independent and identically distributed (i.i.d.) complex circularly-symmetric Gaussian vectors with zero-mean and covariance matrix $M_w$, which is unknown since no prior information as to the interferers number and their channel impulse response is assumed available; similarly, under $H_1$, we model $z(1), \ldots, z(Q)$ as i.i.d. complex circularly-symmetric Gaussian vectors with zero-mean and unknown covariance matrix $M_z$. Finally, we assume that only the spreading code $\beta_0$ of the user to be detected is known, while its data symbols $\{b_0(1), \ldots, b_0(Q)\}$ and its channel-dependent vectors $\{g_0(1), \ldots, g_0(Q)\}$ are regarded as unknown deterministic parameters.

Lacking prior information as to the user parameters and the second-order statistics of the interference, the Neyman-Pearson test [23] cannot be implemented here. A common way to circumvent this drawback is resorting to a GLRT approach which amounts to replacing the unknown quantities with their maximum likelihood (ML) estimates under each hypothesis [23]. Upon defining $W = [w(1) \ldots w(Q)] \in \mathbb{C}^{LMN \times Q}$, $Z = [z(1) \ldots z(Q)] \in \mathbb{C}^{LMN \times Q}$, and $G = [b_0(1)g_0(1), \ldots, b_0(Q)g_0(Q)] \in \mathbb{C}^{D \times Q}$ with $D = (N + 2P)M$, the test (6) can be recast as

$$\begin{cases} H_0 : R = W \\ H_1 : R = C_{0,0} G + Z \end{cases} \quad (7)$$

and the GLRT can be expressed as

$$\frac{\sup_{M_z, G} f_R(R|M_z, G, H_1)}{\sup_{M_w} f_R(R|M_w, H_0)} \underset{H_0}{\overset{H_1}{\gtrless}} \eta, \quad (8)$$

where $\eta$ is a threshold to be set so as to achieve a given probability of false alarm $P_{fa}$ and[2]

$$f_R(R|M_w, H_0) = \underbrace{\frac{c_0}{(|M_w|_p)^H} \exp\left\{-\mathrm{tr}\left\{M_w^+ R R^\dagger\right\}\right\} \delta\left(\Phi_w^\dagger R\right)}_{m_0(R|M_w, H_0)}, \quad (9)$$

$$f_R(R|M_z, G, H_1) = \underbrace{\frac{c_1}{(|M_z|_p)^H} \exp\left\{-\mathrm{tr}\left\{M_z^+ (R - CG)(R - CG)^\dagger\right\}\right\}}_{m_1(R|M_z, G, H_1)} \\ \times \delta\left(\Phi_z^\dagger (R - CG)\right), \quad (10)$$

are the probability density functions (pdf's) of the received data under both hypotheses [24]. In the above expressions, $c_0$ and $c_1$ are positive normalization constants; $\delta(\cdot)$ is the product of the Dirac delta functions of the elements of the matrix argument; finally, $\Phi_x$ is a $LNM \times (LNM - \mathrm{rank}\{M_x\})$ matrix whose columns form an orthonormal basis for the null space of the columns of $M_x$, with $x = w, z$.

Unfortunately, the GLRT (8) cannot be applied in this scenario; in fact, after maximizing the numerator with respect to $M_z$, we have:

$$\sup_{M_z, G} f_R(R|M_z, G, H_1) \\ \propto \sup_{G} \frac{1}{|(R - CG)(R - CG)^\dagger|_p}. \quad (11)$$

[2]For notational simplicity, in the following we drop the subscript in $C_{0,0}$.

Now, the matrix $G$ can always be chosen such that $(R - CG)(R - CG)^\dagger$ has $D$ strictly positive eigenvalues approaching zero, thus causing the right hand side of (11) to diverge. As a result, the GLRT (8) is unbounded, and hence the ML estimate does not exist. This circumstance was also observed in [20, Equation (4)], [25, Equation (5)], and is explained by noticing that the parameter space is too large. A possible solution to this drawback is to resort to the *method of sieves* [19], wherein the parameter space is restricted to a subspace such that the ML estimate exists and is unique. According to the method of sieves, a unique ML solution under $H_1$ exists upon restricting the parameter space to the set

$$\mathcal{S} = \begin{cases} M_z, G | \ \hat{\rho} = \mathrm{rank}(\hat{M}_z) \leq LNM - D \\ M_z = \hat{\Psi}^\dagger \begin{bmatrix} B & O_{\hat{\rho}, 2NM - \hat{\rho}} \\ O_{2NM - \hat{\rho}, \hat{\rho}} & O_{2NM - \hat{\rho}, 2NM - \hat{\rho}} \end{bmatrix} \hat{\Psi} \end{cases},$$

where $\hat{M}_z = (R - CG)(R - CG)^\dagger$, $B$ is an arbitrary $\hat{\rho} \times \hat{\rho}$ matrix and $\hat{\Psi}$ is the matrix of the normalized eigenvectors of $\hat{M}_z$, and we consider the following modified GLRT (MGLRT):

$$\frac{\sup_{(M_z, G) \in \mathcal{S}} m_1(R|M_z, G, H_1)}{\sup_{M_w} m_0(R|M_w, H_0)} \underset{H_0}{\overset{H_1}{\gtrless}} \eta. \quad (12)$$

The next proposition provides a closed-form expression for the test statistic in (12).

*Proposition:* The MGLRT for the binary detection problem in (6) is given by

$$\mathcal{T} = \frac{|RR^\dagger|}{|(I_{LNM} - CC^+)RR^\dagger(I_{LNM} - CC^+)^\dagger|_p} \underset{H_0}{\overset{H_1}{\gtrless}} \eta. \quad (13)$$

*Proof:* Using [20][Lemma 1], it follows that the test (14) can be equivalently recast as

$$\sup_{G : \hat{\rho} \leq LNM - D} \frac{|RR^\dagger|}{|(R - CG)(R - CG)^\dagger|_p} \underset{H_0}{\overset{H_1}{\gtrless}} \eta.$$

Hence, it is sufficient to show that the solution to the problem

$$\hat{G} = \arg \inf_{G : \hat{\rho} \leq LNM - D} |(R - CG)(R - CG)^\dagger|_p$$

is $\hat{G} = C^+ R$. Let $C = U\Sigma V^\dagger$ be the economy-size singular value decomposition (SVD) of $C$, where $U \in \mathbb{C}^{LNM \times D}$ is a tall matrix with orthonormal columns, $\Sigma \in \mathbb{C}^{D \times D}$ is a diagonal matrix containing the non-zero singular values on its main diagonal, and $V \in \mathbb{C}^{D \times D}$ is a unitary matrix. Let $F = [f_1, \ldots, f_D]^H = \Sigma V^\dagger G$. Since $\Sigma V^\dagger$ is invertible, we have that $\hat{G} = V\Sigma^{-1} \hat{F}$ with

$$\hat{F} = \arg \min_{F : \hat{\rho} \leq LNM - D} |(R - UF)(R - UF)^\dagger|_p.$$

Generalizing [25][Proposition 1], it is verified that $\hat{F} = U^\dagger R$. This concludes the proof.

Some remarks are now in order. Notice first that the projector $I_{LNM} - CC^+$ spans the null space of $C$ and, hence, the matrix $(I_{LNM} - CC^+)R$ in the denominator of (13) does not contain the useful signal $CG$. Also, the computation of the test statistic $\mathcal{T}$ can be simplified as outlined in the following. Let $\bar{R} = \bar{U}^\dagger R$, where $\bar{U} \in \mathbb{C}^{LNM \times LNM}$ is a unitary transformation such that

$CC^+ = \bar{U}\bar{\Delta}\bar{U}^\dagger$ and

$$\bar{\Delta} = \begin{bmatrix} O_{LNM-D,LNM-D} & O_{LNM-D,D} \\ O_{D,LNM-D} & I_D \end{bmatrix}.$$

Also, let $[\bar{L}, \mathbf{0}_{LNM,Q-LNM}]\bar{Q}$ be the Q-R decomposition of $\bar{R}$ with $\bar{L} \in \mathbb{C}^{LNM \times LNM}$ a lower triangular matrix and $\bar{Q} \in \mathbb{C}^{Q \times Q}$ a unitary matrix. We have:

$$\begin{aligned}
\mathcal{T} &= \frac{|\bar{R}\bar{R}^\dagger|}{|(I_{LNM} - \bar{\Delta})\bar{R}\bar{R}^\dagger(I_{LNM} - \bar{\Delta})|_p} \\
&= \frac{|\bar{L}\bar{L}^\dagger|}{|(I_{LNM} - \bar{\Delta})\bar{L}\bar{L}^\dagger(I_{LNM} - \bar{\Delta})|_p} \\
&= \prod_{i=LNM-D+1}^{LNM} (\bar{l}_{i,i})^2,
\end{aligned} \quad (14)$$

where $\bar{l}_{i,i}$ denotes the $(i,i)$-th entry of $\bar{L}$. Notice now that computing the Q-R decomposition of $\bar{R}$ only involves $\mathcal{O}(Q(LNM)^2)$ floating point operations [26].

### A. On the detector's CFARness

In the following, we investigate the ability of the proposed detector to operate with a CFAR under the design assumptions. In particular, let $M_w^* \in \mathbb{C}^{LNM \times LNM}$ be the true covariance matrix of the received data under $H_0$ and let $PP^\dagger$ be the Cholesky decomposition of $\bar{U}M_w^*\bar{U}^\dagger$, with $P \in \mathbb{C}^{LNM \times LNM}$ a lower triangular matrix. Then, under $H_0$, we can write $|\bar{R}\bar{R}^\dagger| = |P|^2|VV^\dagger|$ and $(I_{LNM} - \bar{\Delta})\bar{R} = [V_1^\dagger P_1^\dagger, O_{Q,D}]^\dagger$, where $V \in \mathbb{C}^{LNM \times Q}$ is a random matrix whose elements are i.i.d. complex circularly-symmetric Gaussian variables with zero-mean and unit variance, $V_1 \in \mathbb{C}^{(LNM-D) \times Q}$ contains the first $LNM - D$ rows of $V$ and $P_1 \in \mathbb{C}^{(LNM-D) \times (LNM-D)}$ contains the first $LNM - D$ rows and columns of $P$. Finally, letting $[L, O_{LNM,Q-LNM}]Q$ be the Q-R decomposition of $V$, the test statistic in (14) can be recast as

$$\begin{aligned}
\mathcal{T} &= \frac{|P|^2|VV^\dagger|}{|P_1|^2|V_1V_1^\dagger|} \\
&= \underbrace{\prod_{i=LNM-D+1}^{2NM} p_{i,i}^2}_{\mathcal{T}_e} \prod_{j=LNM-D+1}^{2NM} l_{j,j}^2,
\end{aligned} \quad (15)$$

where $p_{i,i}$ and $l_{i,i}$ denote the $(i,i)$-th entry of $P$ and $L$, respectively.

Since $\mathcal{T}_e$ depends upon $P$ (and, hence, upon $M_w^*$), the proposed receiver does not have a CFAR. However, it is possible to make the receiver bounded-CFAR with respect to $M_w^*$ through a suitable normalization of the test statistic. Indeed, let $\mathcal{T}_{e,max}$ be the maximum value of $\mathcal{T}_e$ over all possible scenarios under $H_0$ (which can be obtained via experimental measurements). The normalized test statistic $\mathcal{T}_n = \mathcal{T}/\mathcal{T}_{e,max}$ is upper bounded under $H_0$ by the CFAR test statistic $\mathcal{T}_{\text{CFAR}} = \mathcal{T}/\mathcal{T}_e$. Therefore, if we choose $\eta$ so as $P(\mathcal{T}_{\text{CFAR}} > \eta|H_0) = \bar{P}_{fa}$, then we have

$$P(\mathcal{T}_n > \eta|H_0) \leq P(\mathcal{T}_{\text{CFAR}} > \eta|H_0) = \bar{P}_{fa}. \quad (16)$$

## IV. NUMERICAL RESULTS

The performance of the proposed detection scheme can be expressed in terms of the probability of false alarm $P_{fa}$ and the probability of detection $P_d$. Since no closed form expressions are available for both $P_{fa}$ and $P_d$, we resort here to Monte-Carlo simulations. In the following, we consider a DS/CDMA system with oversampling factor $M = 2$, processing gain $N = 15$, pseudo-noise (PN) spreading codes and bandlimited raised cosine chip waveforms with roll-off factor $\alpha$ and duration $4T_c$. Each user transmits binary data symbols, i.e., $\mathcal{B}_k = \{-1, +1\} \; \forall \; k$, while the processed data vectors $r(1), \ldots, r(Q)$ span $L = 2$ symbol intervals with $Q = 120$. The equivalent baseband channel responses are modeled as 3-path channels with time-varying path gains [13], [14]:

$$c_k(t, \tau) = A_k \sum_{p=1}^{3} \alpha_{k,p}(t)\delta(\tau - \tau_{k,p}), \quad k = 1, \ldots, K. \quad (17)$$

In (17), $A_k$ denotes the received signal amplitude of the $k$-th user; $\{\tau_{k,p}\}$ denote the path delays and are modeled as independent random variables uniformly distributed in $[0, (N-1)T_c]$; finally, $\{\alpha_{k,p}(t)\}$ denote the path gains and are modeled as independent stationary Gaussian processes with zero mean and autocorrelation function set according to the Jakes model [27], i.e. $R(\tau) = \mathrm{E}[\alpha_{k,p}(t)\alpha_{k,p}^*(t-\tau)] = J_0\left(2\pi f_d T_b^{-1}\tau\right)$, where $J_0(\cdot)$ is the zero order Bessel Function of the first kind and $f_d$ is the normalized Doppler frequency shift. In the following, we set $f_d$ to 0.1 or 0.01, which correspond to high user mobility scenarios. We define the Signal-to-Noise-Ratio (SNR) as $QA_0^2/\mathcal{N}_0$; also, we assume $A_1 = \ldots = A_K$ and define the Signal-to-Interference-Ratio (SIR) as $A_0^2/A_1^2$. Unless otherwise stated, we consider a power-controlled scenario with SIR=0 dB and we set the test threshold so as to have $P_{fa} = 10^{-2}$.

In Fig. 1 we plot $P_d$ versus SNR for different values of $K$ and $\alpha = 0.3$. As a benchmark, we also report the detection performance of the genie-GLRT detector which has perfect knowledge of the covariance matrices $M_z$ and $M_w$, i.e.,

$$\frac{\sup_{G} f_R(R|M_z, G, H_1)}{f_R(R|M_w, H_0)} \begin{array}{c} H_1 \\ \gtrless \\ H_0 \end{array} \eta. \quad (18)$$

Taking the logarithm of (18) and after some algebra, we obtain

$$\begin{aligned}
&\sum_{q=1}^{Q} r^\dagger(q)\left(M_w^{-1} - M_z^{-1}\right)r(q) \\
&+ \sum_{q=1}^{Q} r^\dagger(q)M_z^{-1}C\left(C^\dagger M_z^{-1}C\right)^{-1}C^\dagger M_z^{-1}r(q) \begin{array}{c} H_1 \\ \gtrless \\ H_0 \end{array} \eta.
\end{aligned} \quad (19)$$

It is interesting to notice that the detection performance improves for larger $f_d$ since temporal diversity is exploited. Also, the performance loss of the MGLRT with respect to the genie-GLRT is less than 3 dB for $K = 1$ at any SNR, while it becomes larger for $K > 1$ as a consequence of the fact that the estimation accuracy of the unknown covariance matrices under $H_0$ and $H_1$ inevitably degrades. In Fig. 2, instead, we plot $P_d$ versus SNR for different values of the roll-off factor and $K = 3$. As expected, the detection performance improves for increasing $\alpha$ since a larger bandwidth is employed.

So far we have assumed that the user to be detected is active during the entire processing window. However, in real world situations, a new user may start transmitting asynchronously with respect to the beginning of the test. In particular, let $\bar{Q} \leq Q$ and assume that only the data vectors $r(Q - \bar{Q} + 1), \ldots, r(Q)$ contain the useful signal to be detected. Fig. 3 shows that the detection performance gracefully degrades for decreasing values of $\bar{Q}$. This is a very attractive feature, which makes the MGLRT detector suitable for the detection of bursty users.

Finally, in Fig. 4 we remove the power-controlled assumption and we study the detection performance of the MGLRT detector for different values of the SIR. It is seen that $P_d$ is not significantly affected by the power level of the interfering signals. In fact, for negative SIRs the detector suffers no asymptotical loss with respect to the power-controlled scenario, whereas $P_d$ improves for positive SIRs. As a consequence, the MGLRT detector can be effectively employed in near-far scenarios, as for example in neighbor discovery applications wherein no power control can usually be guaranteed among mobiles [7].

## V. Conclusions

In this paper we have considered the problem of detecting the presence of a new user in a DS/CDMA system operating over a doubly-dispersive fading channel. This is of interest in all situations where spread-spectrum signals entering the channel are to be detected as, for example, in cell-search procedures, in handoff management algorithms, and neighbor discovery phases. The proposed detection structure (referred to as MGLRT) is based on the *method of sieves* and is fully blind, meaning that it only requires knowledge of the spreading code of the user to be detected. We have provided an efficient QR-based procedure to evaluate the decision statistic and have also proved that the MGLRT has a bounded CFAR under the design assumptions. The proposed detector achieves satisfactory detection performance in doubly-dispersive channels, even when the user to be detected is only present in a fraction of the processing window and strong cochannel interference is present.

## References

[1] E. Dahlman, P. Beming, J. Knutsson, F. Ovesjö, M. Persson, and C. Roobol, "WCDMA–The radio interface for future mobile multimedia communications," *IEEE Transactions on Vehicular Technology*, vol. 47, no. 4, Nov. 1998.
[2] R. Esmailzadeh, M. Nakagawa, and A. Jones, "TDD-CDMA for the 4th generation of wireless communications," *IEEE Personal Communications*, vol. 10, no. 4, pp. 8–15, 2003.
[3] Y. Sung, Y. Lim, L. Tong, and A.-J. van der Veen, "Signal processing advances for 3G WCDMA: From rake receivers to blind techniques," *IEEE Communications Magazine*, vol. 47, no. 1, pp. 48–54, Jan. 2009.
[4] A. J. Viterbi, *CDMA: Principles of Spread Spectrum Communication*. Addison Wesley, 1995.
[5] Y. Wang and T. Ottosson, "Cell search in W-CDMA," *IEEE Journal on Selected Areas in Communications*, vol. 18, no. 8, pp. 1470–1482, 2000.
[6] D. Angelosante, E. Biglieri, and M. Lops, "Neighbor discovery for wireless networks," in *Proc. IEEE International Symposium on Information Theory*, Nice, France, Jun. 2007.
[7] L. Venturino, X. Wang, and M. Lops, "Multiuser detection for cooperative networks and performance analysis," *IEEE Transactions on Signal Processing*, vol. 54, no. 9, pp. 3315–3329, Sep. 2006.
[8] U. Mitra and H. V. Poor, "Activity detection in a multi-user environment," *Wireless Personal Communications*, vol. 3, no. 1-2, pp. 149–174, 1996.
[9] M. McCloud and L. L. Scharf, "Generalized likelihood detection on multiple access channels," in *Proc. Thirty-First Asilomar Conference on Signals, Systems and Computers*, vol. 2, Pacific Grove, CA, Nov. 1997, pp. 1033–1037.
[10] T. Oskiper and H. Poor, "Activity detection in a spread spectrum network," in *Proc. IEEE Internatinal Symposium on Spread Spectrum Techniques and its Applications*, vol. 1, Parsippany, NJ, Sep. 2000, pp. 310–314.
[11] E. Biglieri and M. Lops, "Multiuser detection in a dynamic environment? Part I: User identification and data detection," *IEEE Transactions on Information Theory*, vol. 53, no. 9, pp. 3158–3170, Sep. 2007.
[12] S. Buzzi, A. D. Maio, and M. Lops, "Code-aided blind adaptive new user detection in DS/CDMA systems with fading time-dispersive channels," *IEEE Transactions on Signal Processing*, vol. 51, no. 10, pp. 2637–264, Oct. 2003.
[13] B. Sklar, "Rayleigh fading channels in mobile digital communication systems – Part I & Part II," *IEEE Communications Magazine*, vol. 35, no. 7, pp. 90–109, Jul. 1997.
[14] J. G. Proakis, *Digital Communications*. Mc-Graw-Hill - fourth edition, 2001.
[15] *EURASIP Journal on Applied Signal Processing:* Special Iusssue on Reliable Communications over Rapidly Time-Varying Channels, 2006.
[16] S. U. Hwang, J. H. Lee, and J. Seo, "Low complexity iterative ICI cancellation and equalization for OFDM systems over doubly selective channels," *IEEE Transactions on Broadcasting*, vol. 55, no. 1, pp. 132–139, Mar. 2009.
[17] S. Lu, B. Narasimhan, and N. Al-Dhahir, "A novel SFBC-OFDM scheme for doubly selective channels," *IEEE Transactions on Vehicular Technology*, vol. 58, no. 5, pp. 2573–2578, Jun. 2009.
[18] S.-W. Hou and C. C. Ko, "Intercarrier interference suppression for OFDMA uplink in time- and frequency-selective fading channels," *IEEE Transactions on Vehicular Technology*, vol. 58, no. 6, pp. 2741–2754, Jul. 2009.
[19] U. Grenander, *Abstract Inference*. New York, NY: John Wiley & Sons, Inc., 1981.
[20] K. Gerlach and M. J. Steiner, "Adaptive detection of range distributed targets," *IEEE Transactions on Signal Processing*, vol. 47, no. 7, pp. 1844–1851, Jul. 1999.
[21] E. Grossi, M. Lops, and L. Venturino, "Linear receivers on MIMO channels: A comparison between DS/CDMA and MC-DS/CDMA," in *Proc. of the International Symposium on Information Theory and its Applications*, Parma, Italy, Oct. 2004.
[22] ——, "Blind schemes for asynchronous CDMA systems on dispersive MIMO channels," *IEEE Transactions on Wireless Communications*, vol. 6, no. 6, pp. 2066–2075, Jun. 2007.
[23] H. L. V. Trees, *Detection, Estimation, and Modulation Theory – Part I, III*. New York, NY: John Wiley & Sons, Inc., 2001.
[24] K. Miller, *Multidimensional Gaussian Distributions*. New York, NY: John Wiley & Sons, Inc., 1964.
[25] A. De Maio, "Polarimetric adaptive detection of range-distributed targets," *IEEE Transactions on Signal Processing*, vol. 50, no. 9, pp. 2152–2159, Sep. 2002.
[26] G. H. Golub and C. F. V. Loan, *Matrix Computations - third edition*. Baltimore, MD: The Johns Hopkins University Press, 1996.
[27] W. C. Jakes, *Microwave Mobile Communications*. Piscataway, NY: IEEE Press, 1994.

6

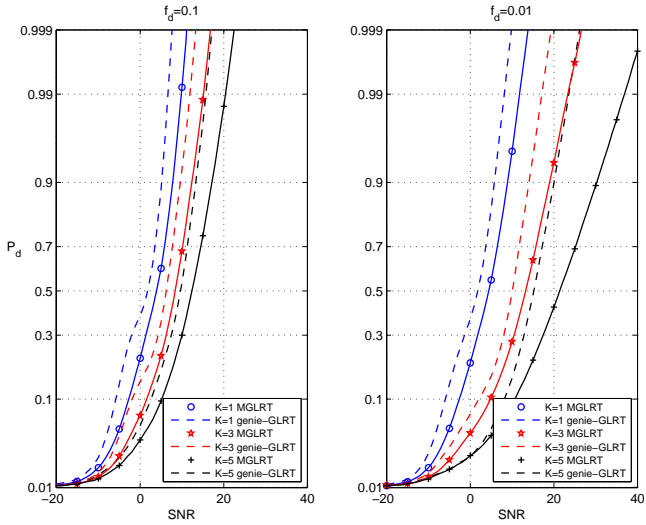

Figure 1. $P_d$ versus SNR for $K = 1, 3, 5$. System parameters: $N = 15$, $M = 2$, $L = 2$, $Q = 120$, $\alpha = 0.3$ and SIR= 0 dB.

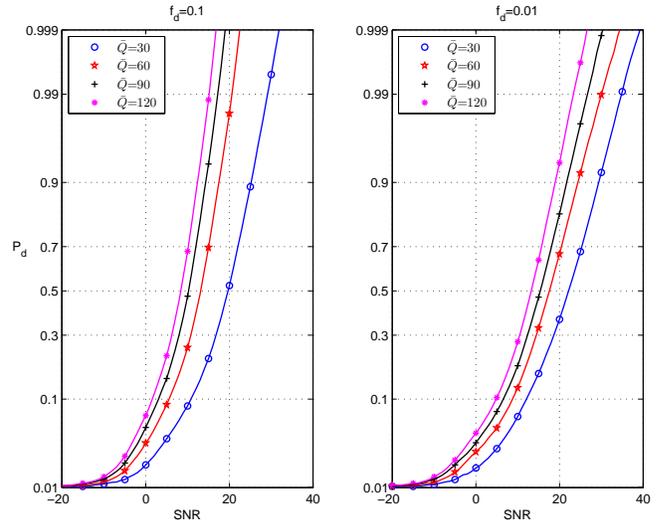

Figure 3. $P_d$ versus SNR for $\bar{Q} = 30, 60, 90, 120$. System parameters: $N = 15$, $M = 2$, $L = 2$, $Q = 120$, $\alpha = 0.3$, $K = 3$ and SIR= 0 dB.

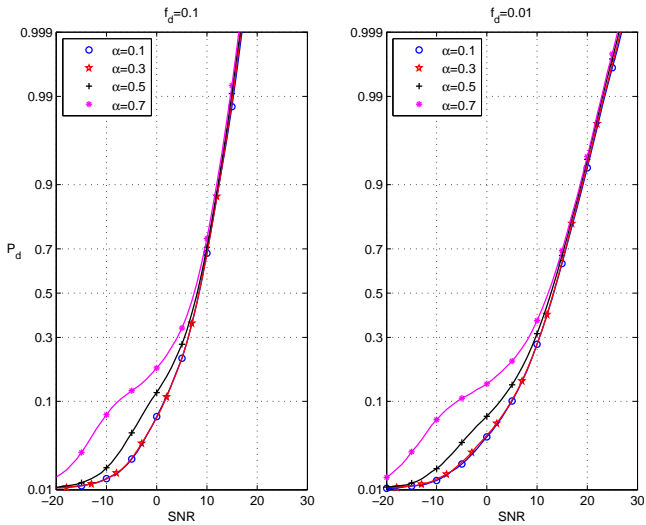

Figure 2. $P_d$ versus SNR for $\alpha = 0.1, 0.3, 0.5, 0.7$. System parameters: $N = 15$, $M = 2$, $L = 2$, $Q = 120$, $K = 3$ and SIR= 0 dB.

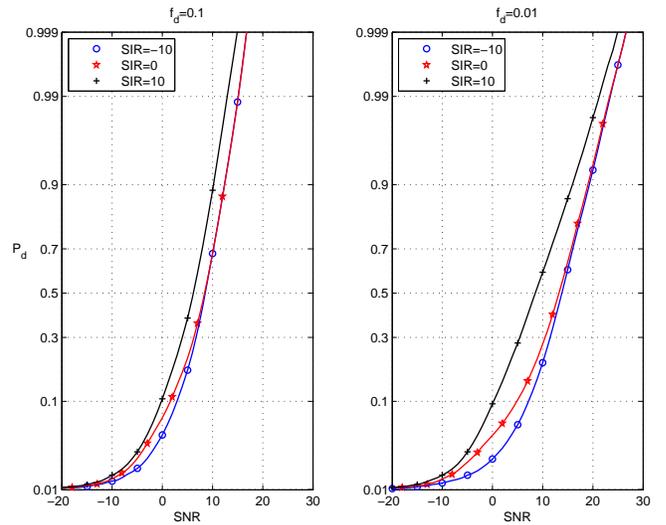

Figure 4. $P_d$ versus SNR for SIR= $-10, 0, 10$ dB. System parameters: $N = 15$, $M = 2$, $L = 2$, $Q = 120$, $\alpha = 0.3$ and $K = 3$.